# Sliding of a liquid spherical drop in an external fluid: a generalization of the Hadamard–Rybczynski equation

Peter Lebedev-Stepanov

Shubnikov Institute of Crystallography, Kurchatov Complex of Crystallography and Photonics, Leninskiy Prospekt 59, Moscow 119333, Russia

E-mail: lebstep.p@crys.ras.ru

Author ORCID https://orcid.org/0000-0002-7009-4319

An analytical solution is obtained for the problem of the slow movement of a small drop of liquid in another immiscible liquid in an infinitely large reservoir with the boundary condition of partial slip at the liquid-liquid interface. That generalizes the conventional Navier condition of partial slip that given at the liquid-solid interface, since the solid (rigid) state can be considered as a liquid with infinite viscosity. A generalized Hadamard–Rybczynski equation (HRE) is obtained. If slip length $\lambda=0$ that equation transforms into the conventional HRE. For infinite viscosity of the droplet, generalized HRE becomes a well-known relation generalizing the Stokes drag force for a solid sphere, taking into account the boundary condition of partial slip. At certain $\lambda$, we arrive at a model with continuity of the all components of the viscous stress tensor at the interface of two fluids, including diagonal tensor components. Generalized HRE is applied to the interpretation of the experiment in which the velocity of falling of a spherical droplet of silicate oil in the castor oil is investigated. Streamlines have been built corresponding to both the Hadamard–Rybczynski model and the partial slip approach. Presumably, the best applicability of the generalized HRE should be expected for the interface of hydrophobic liquid and hydrophilic one (water – hydrocarbons, water – higher alcohols, in general: aqueous emulsions, water – lipophilic organic liquids and oils, etc.). These are quite important emulsions in practical terms, for example, for the oil industry and medicine. Experimental methods for determining the slip length are discussed.

# 1. INTRODUCTION

The Hadamard–Rybczynski equation (HRE) describes the slow motion of a small spherical liquid drop in an external liquid [1-2]. HRE is a generalization of the Stokes equation that corresponds to the motion of a solid sphere. HRE turns into the Stokes equation in the limit of infinite viscosity of the liquid drop [3]. Thus, a solid body is treated as a liquid with infinite viscosity and, accordingly, an infinitely long relaxation time.

Note that both these equations use the no-slip boundary condition [4]. A generalization of the no-slip condition is the partial slip condition (or Navier condition) [5-6]. Currently, it is applied exclusively to the case of a solid-liquid interface [6-10]. Is it possible to extend the application of the partial slip condition to the liquid-liquid interface?

The logic of this question is due to the fact that, as we have just seen, in the problems of fluid mechanics the solid (rigid) state and the liquid differ only in viscosity. Indeed, let us assume that there is an interface of separation between two immiscible liquids, where one liquid slides over the other. If the viscosity of one of the liquids tends to infinity, we obtain a liquid-solid interface with a boundary condition of partial slipping. Thus, it is possible to transform a solid-liquid interface into a liquid-liquid interface by continuously changing the viscosity of one of the liquids. Thus, we obtain a generalization of the Navier boundary condition to the liquid-liquid interface.

The HRE does not always provide a comprehensive description of the experiment [3, 11]. This indicates that the interaction of contacting liquids is not well understood. Various mechanisms are proposed to explain this deviation, for example, increased viscosity at the interface between two liquids and the presence of unaccounted surfactants [3]. But is that consideration inclusive enough?

It is known that the partial slip condition is realized at the hydrophobic-hydrophilic solid-liquid interface [6, 12-14]. Similarly, systems that have a liquid-liquid interface where the partial slip condition is expected to be satisfied are water-oil and oil-water emulsions. There is an interface separating water and a hydrophobic liquid (oil). Obviously, the study of a previously unknown mechanism of interaction of their components has not only fundamental but also practical significance.

This work is devoted to the study of the motion of a spherical drop of liquid in an infinite space filled with another liquid that does not mix with the first liquid. It is assumed that the condition of partial slip (generalized Navier condition) is satisfied at the interface between the two liquids. An equation of motion of a liquid drop is obtained, which is a generalization of the HRE.

Possible applications of the generalized HRE and the possibility of its experimental verification are considered.

## 2. MODEL WITH ARBITRARY SLIP LENGTH AT LIQUID-LIQUID INTERFACE

Let us consider a liquid droplet placed inside another liquid. Both liquids are insoluble in each other, do not mix with each other, and have a clear interface. The drop has a spherical shape stabilized by interfacial surface tension. The $z$ axis is oriented vertically upwards (Fig. 1). If the drop density $\rho'$ is less than the liquid density $\rho$, it floats vertically upwards with a steady-state velocity $V_0$.

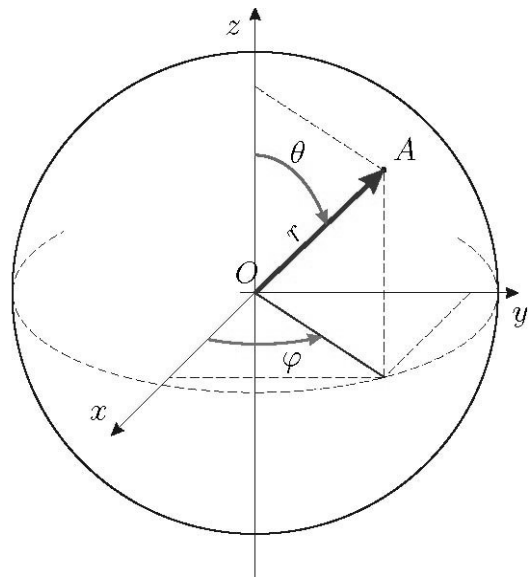

FIG 1. Spherical coordinates $(r, \theta, \varphi)$, which include the radial coordinate, polar and azimuthal angles, respectively.

The hydrostatic pressure in the external liquid on the surface of the drop is determined by the formula

$$p_0 = -\rho g(h - R\cos\theta), \qquad (1)$$

where $h$ is the depth of immersion of the center of the drop. The minus sign before the expression on the right takes into account that the hydrostatic pressure is compressive.

Similarly, the internal liquid creates pressure:

$$p_0' = \rho' g R\cos\theta + C, \qquad (2)$$

where $C$ is a constant.

The surface of the liquid sphere is affected by the resulting pressure equal to the difference between Eq. (2) and Eq.(1)

$$\Delta p_0(\theta) = p_0 - p_0' = (\rho - \rho')gR\cos\theta, \tag{3}$$

where we have neglected the terms independent of the polar angle. It determines the Archimedean force acting on the drop,

$$F_z = 2\pi R^2 \int_0^\pi \Delta p_0(\theta) \sin\theta \cos\theta \, d\theta = (\rho - \rho')gv, \tag{4}$$

where $v = \frac{4}{3}\pi R^3$ is the volume of the liquid sphere.

The constancy of the steady-state velocity of the drop is ensured by the resistance force, which depends on the velocity of the drop, is equal in magnitude to the force (4) and is directed in the opposite direction. The resistance force is associated with the steady-state flow of the external liquid around the drop and the flow of another liquid inside the drop.

In the frame of reference associated with the droplet, the external liquid at $r \to \infty$ satisfies the boundary conditions

$$V_r(\infty, \theta) = V_0 \cos\theta, \tag{5}$$

$$V_\theta(\infty, \theta) = -V_0 \sin\theta. \tag{6}$$

The boundary conditions on the droplet surface are:

$$V_r(R, \theta) = 0, \tag{7}$$

$$V_r'(R, \theta) = 0, \tag{8}$$

which means that the droplet does not change its spherical shape (all the dashed quantities below, like $V_r'$, characterizes the liquid inside the droplet).

In addition, the continuity of viscous stresses acting on the interface between two fluids is used as boundary conditions [3]:

$$\sigma_{rr} = \sigma_{rr}', \tag{9}$$

which in expanded form means

$$-(p + p_0) + 2\eta \frac{\partial V_r}{\partial r} = -(p' + p_0') + 2\eta' \frac{\partial V_r'}{\partial r} \tag{10}$$

and

$$\sigma_{r\theta} = \sigma_{r\theta}', \tag{11}$$

i.e.

$$\eta \left( \frac{1}{r}\frac{\partial V_r}{\partial \theta} + \frac{\partial V_\theta}{\partial r} - \frac{V_\theta}{r} \right) = \eta' \left( \frac{1}{r}\frac{\partial V_r'}{\partial \theta} + \frac{\partial V_\theta'}{\partial r} - \frac{V_\theta'}{r} \right). \tag{12}$$

Here $p$ and $p'$ denote the hydrodynamic pressures in the external and internal liquids, respectively.

Equation (10) can be conveniently rewritten taking into account (1)-(3) as

$$-p + 2\eta \frac{\partial V_r}{\partial r} + p' - 2\eta' \frac{\partial V_r'}{\partial r} = p_0 - p_0' = (\rho - \rho')gR\cos\theta. \tag{13}$$

To describe the motion of liquids at the interface, we introduce the condition of partial slip of the surface of one liquid over the surface of another, which has the form

$$\sigma_{r\theta}(\theta) = \eta \left( \frac{1}{r}\frac{\partial V_r}{\partial \theta} + \frac{\partial V_\theta}{\partial r} - \frac{V_\theta}{r} \right)_{r=R} = \frac{\eta}{\lambda}(V_\theta(R,\theta) - V_\theta'(R,\theta)) \tag{14}$$

and

$$\sigma_{r\theta}'(\theta) = \eta' \left( \frac{1}{r}\frac{\partial V_r'}{\partial \theta} + \frac{\partial V_\theta'}{\partial r} - \frac{V_\theta'}{r} \right)_{r=R} = \frac{\eta'}{\lambda'}(V_\theta'(R,\theta) - V_\theta(R,\theta)), \tag{15}$$

where $\lambda$ and $\lambda'$ are the slip parameters of the external and internal liquids, respectively. Eqs. (14) and (15) generalize the Navier boundary condition of partial slip of a liquid over a solid. Obviously, applying this condition to the liquid-liquid interface, we must use the relative velocity of both liquids on the surface. Taking into account that there is a stitching of the tangential stresses given by Eq. (11), we have

$$\frac{\eta}{\lambda} = -\frac{\eta'}{\lambda'} \quad \text{or} \quad \frac{\lambda}{\lambda'} = -\frac{\eta}{\eta'}. \tag{16}$$

Since conditions (14) and (15) duplicate each other, we will further use only one of them, namely, the condition for the external liquid given by Eq. (14).

Here we are dealing with axisymmetric boundary conditions. Thus, we have to solve an axisymmetric problem. The general solution of the axisymmetric problem for the Stokes equations in a spherical coordinate system is presented in Table 1. The derivation of the formulae is given in Refs. [15-16].

We see that the boundary conditions (5)-(6) and (4) require solutions containing $\cos\theta$ and $\sin\theta$. The solutions for $V_r$ and $p$ have the form of a series expansion in Legendre polynomials, and the solution for $V_\theta$ is represented by a series in the associated Legendre polynomial $P_l^1$. The orthogonality of the polynomials suggests that in order to satisfy the boundary conditions, it is necessary to limit ourselves to the terms of the series that contain $\cos\theta$ or $\sin\theta$. The remaining terms must be discarded, since the boundary conditions do not allow nonzero solutions corresponding to these excess terms to exist. Thus, in the solutions under consideration, we leave only the terms with $l=1$.

Table 1. Solutions of internal and external axisymmetric problems in a spherical coordinate system obtained in the representation of a vector potential. The radial and polar components of the incompressible fluid velocity, pressure and stream function are shown in the rows from top to bottom; $P_l(\cos\theta)$ is the Legendre polynomial, $P_l^1(\cos\theta)$ is the associated Legendre function of the first order.

| Internal problem | External problem |
|---|---|
| $V_r(r,\theta) = \sum_{l=1}^{\infty} l(l+1)\left\{\dfrac{a_l}{4l+6}\left(\dfrac{r}{R}\right)^{l+1} + c_l\left(\dfrac{r}{R}\right)^{l-1}\right\} P_l(\cos\theta)$ | $V_r(r,\theta) = -\sum_{l=1}^{\infty} l(l+1)\left\{\dfrac{b_l}{4l-2}\left(\dfrac{R}{r}\right)^{l} + d_l\left(\dfrac{R}{r}\right)^{l+2}\right\} P_l(\cos\theta) + d_0\left(\dfrac{R}{r}\right)^{2}$ |
| $V_\theta(r,\theta) = \sum_{l=1}^{\infty}\left\{a_l\dfrac{l+3}{4l+6}\left(\dfrac{r}{R}\right)^{l+1} + c_l(l+1)\left(\dfrac{r}{R}\right)^{l-1}\right\} P_l^1(\cos\theta)$ | $V_\theta(r,\theta) = \sum_{l=1}^{\infty}\left\{b_l\dfrac{l-2}{4l-2}\left(\dfrac{R}{r}\right)^{l} + d_l l\left(\dfrac{R}{r}\right)^{l+2}\right\} P_l^1(\cos\theta)$ |
| $p(r,\theta) = \dfrac{\eta}{R}\sum_{l=0}^{\infty}(l+1)a_l\left(\dfrac{r}{R}\right)^{l} P_l(\cos\theta)$ | $p(r,\theta) = -\dfrac{\eta}{R}\sum_{l=1}^{\infty} l b_l\left(\dfrac{R}{r}\right)^{l+1} P_l(\cos\theta)$ |
| $\Psi(r,\theta) = -R^2\sin\theta\sum_{l=1}^{\infty}\left\{\dfrac{a_l}{4l+6}\left(\dfrac{r}{R}\right)^{l+3} + c_l\left(\dfrac{r}{R}\right)^{l+1}\right\} P_l^1(\cos\theta)$ | $\Psi(r,\theta) = R^2\sin\theta\sum_{l=1}^{\infty}\left\{\dfrac{b_l}{4l-2}\left(\dfrac{R}{r}\right)^{l-2} + d_l\left(\dfrac{R}{r}\right)^{l}\right\} P_l^1(\cos\theta) - d_0 R^2\cos\theta$ |

For the internal problem, we have

$$V_r = 2\left\{\dfrac{a}{10}\left(\dfrac{r}{R}\right)^2 + c\right\} P_1, \tag{17}$$

$$V_\theta = \left\{a\dfrac{4}{10}\left(\dfrac{r}{R}\right)^2 + 2c\right\} P_1^1. \tag{18}$$

$$p = p_0 + \dfrac{2a\eta r}{R^2} P_1. \tag{19}$$

For the external problem, we obtain

$$V_r = -2\left\{\dfrac{b}{2}\dfrac{R}{r} + d\left(\dfrac{R}{r}\right)^3\right\}\cos\theta, \tag{20}$$

$$V_\theta = \left\{\dfrac{bR}{2r} - d\left(\dfrac{R}{r}\right)^3\right\}\sin\theta, \tag{21}$$

$$p = -\dfrac{\eta}{R} b\left(\dfrac{R}{r}\right)^2 \cos\theta. \tag{22}$$

The external fluid must satisfy the conditions at infinity. In this case, it is necessary to use the solution of the internal problem given by Eqs. (17)-(19) for $a=0$. Indeed, this will allow us to satisfy conditions (5)-(6). Substituting Eqs. (17)-(19) for $a=0$ into Eqs. (5)-(6) gives

$$2c = V_0. \tag{23}$$

Substituting Eq. (23) in Eqs. (17)-(19), one can obtain

$$V_r = \left\{V_0 - b\frac{R}{r} - 2d\left(\frac{R}{r}\right)^3\right\}\cos\theta, \quad (24)$$

$$V_\theta = \left\{\frac{bR}{2r} - d\left(\frac{R}{r}\right)^3 - V_0\right\}\sin\theta, \quad (25)$$

$$p = -\frac{\eta}{R}b\left(\frac{R}{r}\right)^2\cos\theta. \quad (26)$$

Obviously, the internal fluid must be described only by the solution of the internal problem

$$V_r' = 2\left\{\frac{a'}{10}\left(\frac{r}{R}\right)^2 + c'\right\}\cos\theta, \quad (27)$$

$$V_\theta' = -\left\{a'\frac{4}{10}\left(\frac{r}{R}\right)^2 + 2c'\right\}\sin\theta. \quad (28)$$

$$p' = p_0' + \frac{2a'\eta'r}{R^2}\cos\theta. \quad (29)$$

Substituting Eq. (24) in Eq. (7), we have

$$V_r(R,\theta) = (V_0 - b - 2d)\cos\theta = 0. \quad (30)$$

Hence

$$b = V_0 - 2d. \quad (31)$$

Substituting Eq. (31) into Eqs. (24)-(26), we obtain

$$V_r = \left\{V_0\left(1 - \frac{R}{r}\right) + 2d\left[\frac{R}{r} - \left(\frac{R}{r}\right)^3\right]\right\}\cos\theta, \quad (32)$$

$$V_\theta = \left\{V_0\left(\frac{R}{2r} - 1\right) - d\left[\frac{R}{r} + \left(\frac{R}{r}\right)^3\right]\right\}\sin\theta, \quad (33)$$

$$p = -\frac{\eta}{R}(V_0 - 2d)\left(\frac{R}{r}\right)^2\cos\theta. \quad (34)$$

Similarly, taking into account the condition (8), for the radial component of the velocity of the internal fluid (27) we have

$$V_r'(R,\theta) = 2\left\{\frac{a'}{10} + c'\right\}\cos\theta = 0. \quad (35)$$

Hence

$$a' = -10c'. \quad (36)$$

Then, the system of equations for the internal fluid (27)-(29) can be rewritten as

$$V_r' = 2c'\left\{1-\left(\frac{r}{R}\right)^2\right\}\cos\theta, \tag{37}$$

$$V_\theta' = 2c'\left\{2\left(\frac{r}{R}\right)^2-1\right\}\sin\theta, \tag{38}$$

$$p' = -\frac{20c'\eta' r}{R^2}\cos\theta. \tag{39}$$

Conditions given by Eqs. (7)-(8) imposed in Eq. (32) and (37) on the surface of the drop give

$$\left.\frac{\partial V_r}{\partial \theta}\right|_{r=R} = \left.\frac{\partial V_r'}{\partial \theta}\right|_{r=R} = 0. \tag{40}$$

Therefore, the condition of stitching stresses $\sigma_{r\theta}$ giving by Eq. (12) is simplified

$$\eta\left(\left.\frac{\partial V_\theta}{\partial r}\right|_{r=R} - \frac{V_\theta(R)}{R}\right) = \eta'\left(\left.\frac{\partial V_\theta'}{\partial r}\right|_{r=R} - \frac{V_\theta'(R)}{R}\right). \tag{41}$$

Next, we have

$$V_\theta(R) = -\left\{\frac{1}{2}V_0 + 2d\right\}\sin\theta, \tag{42}$$

$$\frac{\partial V_\theta}{\partial r} = \left\{-V_0\left(\frac{R}{2r^2}\right) + d\left[\frac{R}{r^2} + \frac{3}{R}\left(\frac{R}{r}\right)^4\right]\right\}\sin\theta, \tag{43}$$

$$\left.\frac{\partial V_\theta}{\partial r}\right|_{r=R} = \frac{8d - V_0}{2R}\sin\theta, \tag{44}$$

$$\left.\frac{\partial V_r}{\partial \theta}\right|_{r=R} - \frac{V_\theta(R)}{R} = \frac{8d - V_0}{2R}\sin\theta + \frac{1}{R}\left\{\frac{1}{2}V_0 + 2d\right\}\sin\theta = \frac{6d}{R}\sin\theta. \tag{45}$$

Similarly,

$$V_\theta'(R) = 2c'\sin\theta, \tag{46}$$

$$\frac{\partial V_\theta'}{\partial r} = 2c'\frac{1}{R}\left\{4\frac{r}{R}\right\}\sin\theta, \tag{47}$$

$$\left.\frac{\partial V_\theta}{\partial r}\right|_{r=R} = 8c'\frac{1}{R}\sin\theta, \tag{48}$$

$$\left.\frac{\partial V_\theta'}{\partial r}\right|_{r=R} - \frac{V_\theta'(R)}{R} = \frac{6c'}{R}\sin\theta, \tag{49}$$

Substituting Eq. (45) and Eq. (49) into Eq. (41), we obtain

$$c' = \frac{\eta}{\eta'}d. \tag{50}$$

Then, the Eqs. (37)-(39) can be rewritten as

$$V_r' = 2\frac{\eta}{\eta'}d\left\{1-\left(\frac{r}{R}\right)^2\right\}\cos\theta, \quad (51)$$

$$V_\theta' = 2\frac{\eta}{\eta'}d\left\{2\left(\frac{r}{R}\right)^2 - 1\right\}\sin\theta. \quad (52)$$

$$p' = -\frac{\eta}{\eta'}d\frac{20\eta'r}{R^2}\cos\theta = -d\frac{20\eta r}{R^2}\cos\theta. \quad (53)$$

Let us calculate the expression, taking into account Eqs. (32) and (34)

$$-p(R) + 2\eta\left.\frac{\partial V_r}{\partial r}\right|_R = \frac{3\eta}{R}(V_0 + 2d)\cos\theta. \quad (54)$$

Similarly, we find

$$-p'(R) + 2\eta'\left.\frac{\partial V_r'}{\partial r}\right|_{r=R} = d\frac{20\eta r}{R^2}\cos\theta - \frac{8\eta'\eta}{\eta'R^2}d\cos\theta = \frac{12d\eta}{R}\cos\theta. \quad (55)$$

Substituting Eq. (54) and Eq. (55) into the boundary condition Eq. (13), we obtain

$$V_0 = 2d + \frac{(\rho-\rho')gR^2}{3\eta} \quad (56)$$

The last boundary condition to be taken into account is the partial slip condition for the liquid-liquid interface (14). Substituting Eq. (40) in Eq. (14), we have:

$$\lambda\left(\left.\frac{\partial V_\theta}{\partial r}\right|_{r=R} - \frac{V_\theta(R,\theta)}{R}\right) = V_\theta(R,\theta) - V_\theta'(R,\theta). \quad (57)$$

Substituting into Eq. (57) the projections of the velocities $V_\theta$ and $V_\theta'$, determined by Eqs. (33) and (52), respectively, as well as Eq. (45), we find an equation that allows us to calculate the coefficient $d$:

$$\lambda\frac{6d}{R}\sin\theta = \left(-\left\{\frac{1}{2}V_0 + 2d\right\} - 2\frac{\eta}{\eta'}d\right)\sin\theta, \quad (58)$$

or

$$d = -\frac{(\rho-\rho')gR^2}{6\eta}\left(\frac{6\lambda}{R} + 3 + \frac{2\eta}{\eta'}\right)^{-1}. \quad (59)$$

Substituting Eq. (59) into Eq. (56), one can obtain

$$V_0 = \frac{(\rho-\rho')gR^2}{3\eta}\left\{1 - \left(\frac{6\lambda}{R} + 3 + \frac{2\eta}{\eta'}\right)^{-1}\right\} \quad (60)$$

The remaining undefined coefficients (31), (50) and (36) take the form

$$b = \frac{(\rho - \rho')gR^2}{3\eta}, \tag{61}$$

$$c' = -\frac{(\rho - \rho')gR^2}{6\eta'}\left(\frac{6\lambda}{R} + 3 + \frac{2\eta}{\eta'}\right)^{-1}, \tag{62}$$

$$a' = \frac{5(\rho - \rho')gR^2}{3\eta'}\left(\frac{6\lambda}{R} + 3 + \frac{2\eta}{\eta'}\right)^{-1}. \tag{63}$$

Let us write out the solutions for the external and internal liquids by substituting Eq. (59) in Eqs. (32)-(34) and Eqs. (51)-(53), respectively.

1) The external liquid

$$V_r = \frac{(\rho - \rho')gR^2}{3\eta}\left\{1 - \frac{R}{r} - \left(\frac{6\lambda}{R} + 3 + \frac{2\eta}{\eta'}\right)^{-1}\left[1 - \left(\frac{R}{r}\right)^3\right]\right\}\cos\theta, \tag{64}$$

$$V_\theta = \frac{(\rho - \rho')gR^2}{6\eta}\left\{\frac{R}{r} - 2 + \left(\frac{6\lambda}{R} + 3 + \frac{2\eta}{\eta'}\right)^{-1}\left[2 + \left(\frac{R}{r}\right)^3\right]\right\}\sin\theta, \tag{65}$$

$$p = -\frac{(\rho - \rho')gR^3}{3r^2}\cos\theta. \tag{66}$$

Stream function is (Table 1):

$$\Psi(r,\theta) = \frac{(\rho - \rho')gR^4}{6\eta}\sin^2\theta\left\{\left(\frac{r}{R}\right)^2 - \frac{r}{R} + \left(\frac{6\lambda}{R} + 3 + \frac{2\eta}{\eta'}\right)^{-1}\left(\frac{R}{r} - \left(\frac{r}{R}\right)^2\right)\right\}. \tag{67}$$

2) The internal liquid

$$V_r' = -\frac{(\rho - \rho')gR^2}{3\eta'}\left(\frac{6\lambda}{R} + 3 + \frac{2\eta}{\eta'}\right)^{-1}\left\{1 - \left(\frac{r}{R}\right)^2\right\}\cos\theta, \tag{68}$$

$$V_\theta' = -\frac{(\rho - \rho')gR^2}{3\eta'}\left(\frac{6\lambda}{R} + 3 + \frac{2\eta}{\eta'}\right)^{-1}\left\{2\left(\frac{r}{R}\right)^2 - 1\right\}\sin\theta, \tag{69}$$

$$p' = \frac{10(\rho - \rho')gr}{3}\left(\frac{6\lambda}{R} + 3 + \frac{2\eta}{\eta'}\right)^{-1}\cos\theta. \tag{70}$$

Stream function is (Table 1):

$$\Psi' = \frac{(\rho - \rho')gR^4}{6\eta'}\sin^2\theta\left(\frac{6\lambda}{R} + 3 + \frac{2\eta}{\eta'}\right)^{-1}\left\{\left(\frac{r}{R}\right)^4 - \left(\frac{r}{R}\right)^2\right\}. \tag{71}$$

## 5. DRAG COEFFICIENT $C_D$

Eq. (60) describes the steady-state velocity of a spherical liquid drop in an external fluid. It can also be represented in the equivalent form

$$V_0 = \frac{2(\rho-\rho')gR^2}{3\eta} \frac{1+\eta\eta'^{-1}+3\lambda R^{-1}}{3+2\eta\eta'^{-1}+6\lambda R^{-1}} \tag{72}$$

or

$$V_0 = \frac{2(\rho-\rho')gR^2}{3\eta} \frac{\eta+\eta'(1+3\lambda R^{-1})}{2\eta+3\eta'(1+2\lambda R^{-1})}, \tag{73}$$

or

$$V_0 = \frac{2(\rho-\rho')gR^2}{3\eta} \frac{\alpha+1+3\lambda R^{-1}}{2\alpha+3+6\lambda R^{-1}}, \tag{74}$$

where $\alpha = \dfrac{\eta}{\eta'}$ is the viscosity ratio. Note that (74) can also be represented as

$$V_0 = \frac{(\rho-\rho')gR^2}{3\eta}\left\{1-\left(\frac{6\lambda}{R}+3+2\alpha\right)^{-1}\right\} \tag{75}$$

At $\lambda=0$, we have the no-slip condition, under which Eq. (73) becomes the Hadamard–Rybczynski equation (HRE):

$$V_0 = \frac{2(\rho-\rho')gR^2}{3\eta} \frac{\eta+\eta'}{2\eta+3\eta'}. \tag{76}$$

At $\lambda=0$, condition given by Eq. (57) takes the form

$$V_\theta(R,\theta)-V_\theta'(R,\theta) = 0. \tag{77}$$

It is no-slip condition that corresponds to the Hadamard–Rybczynski model. Thus, the resulting Eq. (60) generalizes the Hadamard–Rybczynski equation and its limit is the Stokes formula by introducing partial slip.

If $\alpha \to 0$, i.e. the viscosity of the drop is much greater than the viscosity of the surrounding liquid, and $\lambda$ has a value limited in absolute value from above, Eq. (74) takes the form

$$V_0 = \frac{2(\rho-\rho')gR^2}{9\eta} \frac{1+3\lambda R^{-1}}{1+2\lambda R^{-1}}, \tag{78}$$

Eq. (78) describes the sliding of a solid sphere in an external fluid with viscosity $\eta$ and slip length $\lambda$. It coincides with the previously published result [4,9]. At $\lambda=0$, Eq. (78) becomes the Stokes equation that describes the motion of a solid sphere under the no-slip boundary condition:

$$V_0 = \frac{2(\rho-\rho')gR^2}{9\eta}. \tag{79}$$

Note that, taking into account Eq. (16), we have

$$\alpha = -\frac{\lambda}{\lambda'}. \tag{80}$$

Then, the original formula (74) transforms to the Stokes equation in the limit $\lambda \to 0$, when simultaneously $\lambda R^{-1} \to 0$ and $\lambda(\lambda')^{-1} \to 0$, i.e., when the external liquid satisfies to a condition close to no-slip, but at the same time, the partial slip regime must take place in the internal liquid. This means that the Stokes limit (79) can be fulfilled not only for a solid sphere moving in the no-slip mode, as was previously assumed, but also under the condition of partial slip of the internal liquid, which is assumed to be much more viscous than the external one.

Another extreme case is when the viscosity of the external liquid is much higher than the viscosity of the drop. If $\lambda$ is a limited quantity, the substitution $\alpha \to \infty$ transforms Eq. (74) into the following expression

$$V_0 = \frac{(\rho - \rho')gR^2}{3\eta}. \tag{81}$$

The HRE has the same limit.

Note that Eq. (74) gives another way to obtain Eq. (81), which is absent in the HRE: i.e., let $\lambda R^{-1} \to \infty$ and, at the same time, $\alpha$ is a limited value. This means that the slip of the external liquid is almost complete, and that of the internal liquid is the same, taking into account Eq. (80).

Considering that the force acting on a liquid drop in an external liquid is determined by the formula

$$F_0 = \frac{4\pi}{3}(\rho - \rho')gR^3. \tag{82}$$

Then the friction coefficient is determined by the formula

$$\beta = \frac{F_0}{V_0} = 2\pi R\eta \frac{2\alpha + 3 + 6\lambda R^{-1}}{\alpha + 1 + 3\lambda R^{-1}}. \tag{83}$$

In the limit $\lambda R^{-1} \to 0$ and $\alpha \to 0$ that corresponds to a solid body with the boundary condition no-slip, Eq. (83) takes the form $\beta = 6\pi R\eta$, which coincides with the Stokes friction coefficient, as it should be. In the case of complete slip, i.e., $\lambda R^{-1} \to \infty$, on can obtain $\beta = 4\pi R\eta$.

We now obtain an expression for the frequently used drag coefficient $C_D$ [17] that is determined according to the relation

$$C_D = \frac{2F_0}{\pi R^2 \rho V_0^2} = \frac{2\beta}{\pi R^2 \rho V_0}. \tag{84}$$

Using Eq. (83) and the Reynolds number

$$\mathrm{Re} = \frac{2\rho R V_0}{\eta}, \tag{85}$$

rewrite Eq. (84) as

$$C_D = \frac{4\beta}{\pi R \eta \,\mathrm{Re}} = \frac{8}{\mathrm{Re}} \frac{2\alpha + 3 + 6\lambda R^{-1}}{\alpha + 1 + 3\lambda R^{-1}}. \tag{86}$$

In the Hadamard–Rybczynski limit, i.e., if $\lambda R^{-1} \to 0$, we obtain the well-known expression [18]

$$C_{D,AR} = \frac{8}{\mathrm{Re}} \frac{2\alpha + 3}{\alpha + 1}. \tag{87}$$

In the case of complete slip, i.e., if $\lambda R^{-1} \to \infty$, one can obtain

$$C_D = \frac{16}{\mathrm{Re}}. \tag{88}$$

## 6. MODEL WITH CONTINUOUS VISCOUS STRESS TENSOR

In the framework of the axisymmetric problem considered above, the continuities of only two non-zero components of the viscous stress tensor were ensured, $\sigma_{rr}$ and $\sigma_{r\theta}$, according to conditions that are given by Eqs. (9) and (11). But there are two more non-zero stresses that do not vanish in the axisymmetric problem, which act tangentially to the spherical surface of the droplet [19,20], namely:

$$\sigma_{\theta\theta} = -p - p_0 + 2\eta \left( \frac{1}{r} \frac{\partial V_\theta}{\partial \theta} + \frac{V_r}{r} \right), \tag{89}$$

$$\sigma_{\varphi\varphi} = -p - p_0 + 2\eta \left( \frac{V_r}{r} + \frac{V_\theta \cot \theta}{r} \right). \tag{90}$$

and similar tensions in the liquid inside the drop.

In general, if these stress tensor components in the external and internal liquids are not stitched together, instability and turbulence will arise. Indeed, let us imagine that the interface between two liquids experiences small oscillations, i.e. surface waves. In this case, local deviations of the normal to the area from the radial direction occur, i.e. non-zero projections in the direction of the polar and axial axes, so that forces corresponding to stresses given by Eqs. (89) and (90) are transmitted to the interface, which can amplify surface waves, creating a turbulent non-stationary picture, the description of which goes beyond the stationary model under consideration.

So, let us consider the stitching conditions:

$$\sigma_{\theta\theta} = \sigma_{\theta\theta}', \tag{91}$$

$$\sigma_{\varphi\varphi} = \sigma_{\varphi\varphi}'. \tag{92}$$

The slip length is calculated in the Appendix, which ensures the stitching of all the above components of the viscous stress tensor. The result looks like this. The sliding of the external fluid along the interface is described by the slip length

$$\lambda = \frac{R}{3}\left(1 - \frac{\eta}{\eta'}\right), \tag{93}$$

whereas for the internal fluid we have a slip length of the form

$$\lambda' = \frac{R}{3}\left(1 - \frac{\eta'}{\eta}\right). \tag{94}$$

The velocity of movement of a drop is calculated in the Appendix:

$$V_0 = \frac{4(\rho - \rho')gR^2}{15\eta}. \tag{95}$$

Obviously, Eqs. (93) and (94) satisfy the condition given by Eq. (16). If the viscosities of both liquids are equal, $\eta = \eta'$, then the slip length becomes zero, and a boundary condition is no-slip. In this case, Eqs. (96) and (74) become identical.

Note that if $\eta > \eta'$, then Eq. (93) corresponds to $\lambda < 0$. Thus, the slip length in the framework of the considered model may be negative.

## 7. DISCUSSION

The HRE is usually applied to describe the motion of a small droplet of liquid in another liquid. In its derivation, the no-slip boundary condition at the liquid-liquid interface, Eq.(77), was used. This somewhat contradicts the initial assumption that both liquids are immiscible (poorly soluble in each other). Therefore, a more natural and generalizing condition is the partial slip of one liquid on the surface of the other one given by Eqs. (14)-(15), which is a modification of the Navier condition originally introduced for the liquid-solid interface.

In this paper, the Navier condition is generalized by its application to the liquid-liquid interface for the first time. A generalized HRE given by Eq.(74) is obtained that transforms into the usual HRE (76) at λ=0.

At λ defined by Eq. (93), we arrive at a model with a slip length controlled by the continuity of the components of the viscous stress tensor at the interface of two fluids. For infinite viscosity of the drop, Eq. (74) becomes a well-known relation generalizing the Stokes drag force for a solid sphere, taking into account the boundary condition of partial slip Eq. (78), obtained using the Navier boundary condition [4,9].

The equation (74), which can be rewritten as

$$V_0 = \frac{2(\rho - \rho')gR^2}{3\eta} \frac{\eta + \eta' + 3\lambda\eta' R^{-1}}{2\eta + 3\eta' + 6\lambda\eta' R^{-1}}. \tag{96}$$

of course, does not exhaust all possible variants of boundary conditions, and therefore cannot claim universality, but there is a fairly wide and important class of substances for the description of which this equation can be applied. From a mathematical point of view, it is obvious that this is approximately the range of substances for the description of which the Boussinesq equation can be used [3,21]:

$$V_0 = \frac{2(\rho-\rho')gR^2}{3\eta} \frac{\eta+\eta'+2e(3R)^{-1}}{2\eta+3\eta'+2eR^{-1}}, \qquad (97)$$

where $e$ is the "coefficient of surface viscosity" introduced by Boussinesq. Indeed, the structure of expression (97) almost exactly coincides with the expression (96) we obtained. In this case, the concept of the slip length seems to make more physical sense.

There are numerous experiments [3] that demonstrate the influence of surfactants on the rate of fall or rise of liquid droplets. These experiments can obviously be naturally understood taking into account that surfactants undoubtedly affect the slip length.

Note that at high coefficients of surface viscosity, $e \to \infty$, Eq. (97) takes the form given by Eq. (79), i.e. it turns into the Stokes drag force formula for a solid sphere. The physical meaning of this is clear: the drop becomes solid at the relaxation times that correspond to the extra-viscous drop surface. In the same way, the HRE (76) turns into the Stokes Eq. (79) for infinitely high viscosity droplets. In the case of Eq. (97), it is sufficient to assume that the surface of a liquid drop has infinite viscosity.

A different asymptotic behavior occurs in the obtained formula (96) in the limit $\lambda \to \infty$: in this case, Eq. (96) takes the form given by Eq. (81).

An infinitely long slip length is the absolute slip mode at the liquid-liquid interface. The HRE (76) has the same limit, but at $\eta \to \infty$, i.e., when the viscosity of the medium is many times greater than the viscosity of the droplet moving in it. This is physically a completely different condition. Note that Eq. (83) takes on the same form for finite slip lengths at $\eta \to \infty$, as it should.

The conditions of applicability of the obtained generalization of the HRE are the same as for the proposed partial slip model. First of all, this is the condition $\text{Re} \ll 1$ under which the linearized Navier-Stokes equations are valid. For raising or falling droplets, the velocity of which is determined by Eq. (74), this condition limits the maximum size of the droplet and, taking into account Eq. (85), has the form

$$R^3 \ll \frac{\eta^2}{\rho|\rho-\rho'|g} \frac{2\alpha+3+6\lambda R^{-1}}{\alpha+1+3\lambda R^{-1}}. \qquad (98)$$

Reducing the inequality by extracting the cubic root, we obtain

$$R < \left(\frac{\eta^2}{\rho|\rho-\rho'|g} \frac{2\alpha+3+6\lambda R^{-1}}{\alpha+1+3\lambda R^{-1}}\right)^{\frac{1}{3}} \qquad (99)$$

Given that the second fraction in parentheses in Eq. (99), as a rule, has a value of the order of 1, the expression can be used to roughly estimate the drop radius

$$R < \left(\frac{\eta^2}{\rho|\rho-\rho'|g}\right)^{\frac{1}{3}} \qquad (100)$$

At typical values of parameters $\eta \sim 10^{-3}\,\text{Pa}\cdot\text{s}$, $\rho \sim 10^3\,\text{kg}\cdot\text{m}^{-3}$, $|\rho-\rho'| \sim 100\,\text{kg}\cdot\text{m}^{-3}$, we have a limitation $R < 10^{-4}\,\text{m}$, i.e., it is an emulsion containing very small droplets with radius of 100 microns or less. These can be, for example, oil-in-water or water-in-oil emulsions. In such systems with a hydrophobic-hydrophilic liquid-liquid interface, partial slip of liquid particles should be expected in the dispersed phase. The study of electrophoresis or sedimentation processes in such emulsions can provide information about the value of slip length $\lambda$.

If highly viscous liquids with viscosity $\eta \sim 1\,\text{Pa}\cdot\text{s}$, such as glycerin or castor oil, are used as the external liquid, then according to Eq. (100), within the framework of this concept, it is permissible to consider the motion of drops several mm in size.

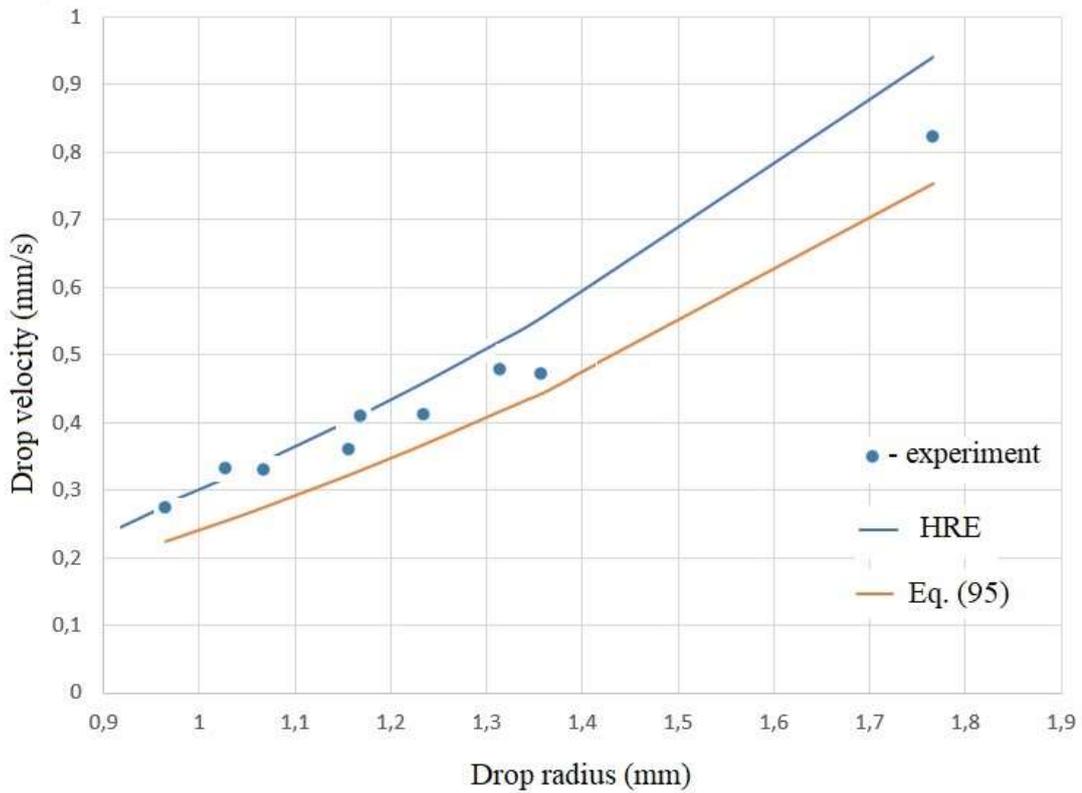

FIG. 2. The velocity of the fall of a spherical drop of silicate oil in castor oil. The blue curve corresponds to the HRE given by Eq. (76). The orange curve visualizes Eq. (95) that corresponds to partial slip on the liquids-liquid interface. The dots adapt the experiment presented in Ref. [22].

The falling of a spherical drop of silicate oil ($\rho' = 1023\,\text{kg}\cdot\text{m}^{-3}$, $\eta' = 0.0232\,\text{Pa}\cdot\text{s}$) in the castor oil ($\rho = 958\,\text{kg}\cdot\text{m}^{-3}$, $\eta = 0.693\,\text{Pa}\cdot\text{s}$) was experimentally studied [22]. The experimental results in the form of dots corresponding to different radii of the drops are adapted in Fig. 2. The experimental dots are located between two theoretical curves: the upper one visualizes the HRE, and the lower one corresponds to the version of partial slip model, given by Eq. (95), assuming the continuity of all components of the viscous stress tensor.

Fig. 3 shows the graphs of the functions $V_\theta'(r, \frac{\pi}{2})$ and $V_\theta(r, \frac{\pi}{2})$ that describe the dependence of the θ-component of the liquid velocity on the radial coordinate $r$ at the polar angle $\theta = \frac{\pi}{2}$ in the system of liquids corresponding to Fig. 2. $V_\theta'(r, \frac{\pi}{2})$ gives the velocity inside the liquid droplet, and $V_\theta(r, \frac{\pi}{2})$ is the velocity outside it. The graphs are plotted for a droplet with a radius of $R$=1 mm.

Fig. 3(a) describes the velocity profile according to the Hadamard–Rybczynski model, corresponding to the blue curve in Fig. 2. The following equations are used to build the graph: Eq. (69) is used for the internal liquid (r<1), Eq. (65) – for the external one (r>1). Zero slip length ($\lambda = 0$) and $\theta = \frac{\pi}{2}$ have to be substituted into Eqs. (65) and (69). Note that the θ-component of the liquid velocity does not change at the liquid-liquid interface, i.e. at $r$=1.

On the other side, Fig. 3(b) describes the θ-component of the liquid velocity inside and outside the droplet at a non-zero slip length that corresponds to the orange curve in Fig. 2. Within the model corresponding to the orange curve, the slip length is controlled by the continuity condition of all components of the viscous stress tensor. For the internal liquid, slip length λ is determined by Eq. (94), and for the external liquid, by formula (93). Substituting the parameters of the system under consideration yields: $\lambda' = -0.159$ mm and $\lambda = 0.108$ mm. Substituting $\lambda'$ into Eq. (69) for r<1, and λ into (65) for r>1 yields the velocity profile shown in Fig. 3(b). Note that instead of Eqs. (69) and (65), one could use formulae (50) and (45), respectively, where the slip lengths given by Eqs. (94) and (93) have already been substituted.

Comparing Fig. 3(a) and (b), one can notice that the θ-components of the velocity of the external liquid in both cases have the same order of magnitude, while the velocities of the internal liquid differ by an order of magnitude. In the considered model with a non-zero slip length, the θ-component of the velocity has a large jump at the droplet interface, but the direction of tangential liquid motion on both sides of the interface is the same.

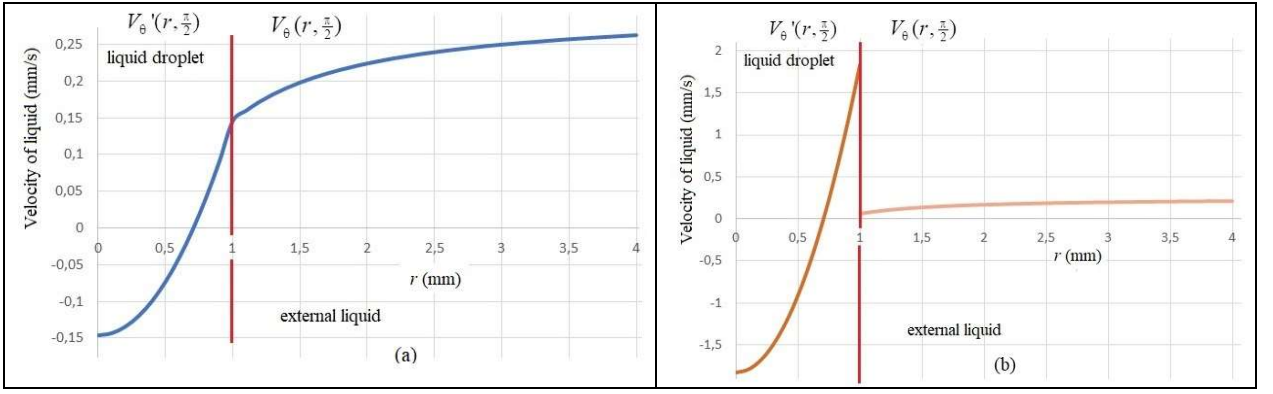

FIG.3. Polar component of the velocity of the liquid inside the drop with a radius of 1 mm (red line), $V_\theta'(r, \frac{\pi}{2})$, and outside of the drop, $V_\theta(r, \frac{\pi}{2})$, given by: (a) – Hadamard–Rybczynski model; (b) – partial slip model, Eqs. (A45) and (A50).

Fig. 4 shows the fluid streamlines corresponding to the cases under consideration.

To construct the streamlines using the Hadamard–Rybczynski model, we used the stream functions given by Eqs. (67) and (71) for the external liquid (castor oil) and internal one (silicate oil), i.e. into the spherical droplet, respectively, in which the no-slip condition $\lambda = 0$ was accepted. Similarly, to construct the streamlines using the partial slip model, we used Eqs. (A48) and (A53) (see the Appendix) for the external liquid and the droplet, respectively.

In both cases, the stream functions were normalized by dividing them by the factor

$$A = \frac{(\rho' - \rho)gR^4}{30\eta}, \tag{103}$$

so that the lines were constructed for a function of the form

$$\psi = A^{-1}\Psi. \tag{104}$$

Thus, substituting the parameters of the liquids under consideration (castor and silicate oils), we obtain the following functions to construct the streamlines. Within the Hadamard–Rybczynski model, we have

$$\psi(r,\theta) \approx -5\sin^2\theta \left\{ \left(\frac{r}{R}\right)^2 - \frac{r}{R} + 0.0159\left(\frac{R}{r} - \left(\frac{r}{R}\right)^2\right) \right\} \tag{105}$$

for the external liquid, and

$$\psi' \approx -2.38\sin^2\theta \left\{ \left(\frac{r}{R}\right)^4 - \left(\frac{r}{R}\right)^2 \right\}. \tag{106}$$

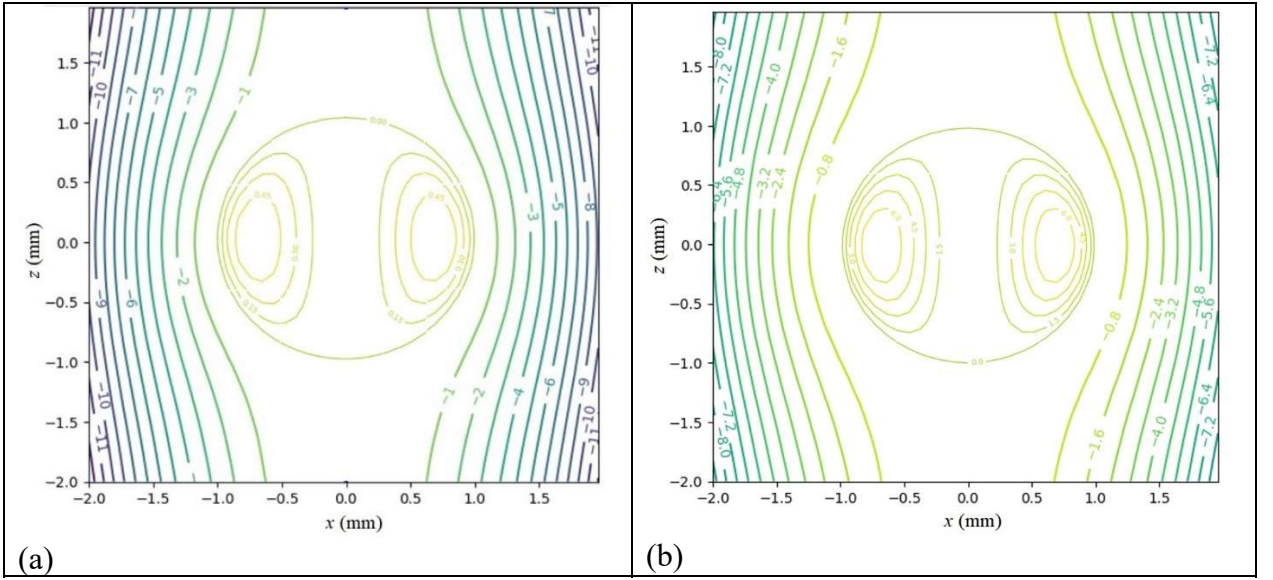

(a)                                                                     (b)

FIG.4. The falling of a spherical drop of silicate oil with a radius of 1 mm in the castor oil: (a) streamlines corresponding to the Hadamard–Rybczynski model; (b) the partial slip approach.

for the spherical droplet (i.e. the internal liquid). In the partial slip model with a continuous viscous stress tensor, we have

$$\psi = -\sin^2\theta \left\{ \frac{R}{r} - \frac{5r}{R} + 4\left(\frac{r}{R}\right)^2 \right\} \quad (107)$$

and

$$\psi' \approx -29.9 \sin^2\theta \left\{ \left(\frac{r}{R}\right)^4 - \left(\frac{r}{R}\right)^2 \right\}. \quad (108)$$

for the external liquid and the droplet, respectively.

      Fig. 4 confirms the conclusion made earlier on the basis of Fig. 3: the main difference between the predictions of the two models under consideration is the magnitude of the velocity of the toroidal flows in the falling droplet: the partial slip model (b) predicts that this velocity should be an order of magnitude greater than in the Hadamard–Rybczynski model (a). Thus, an experimental study of the liquid velocity inside the droplet could show which model is closer to reality.

      Note that in the previous example, a particular type of model with partial slip was considered that gives a good approximation to the experiment shown in Fig. 2. In this case, the slip length is calculated based on the viscosity coefficients and the drop radius given by Eq. (93). This corresponds to a model with continuous components of the viscous stress tensor (Appendix).

      However, in the general case, the continuity of the viscous stress tensor may not be satisfied, so that the slip length $\lambda$ can be considered as an independent parameter. Then, the

question of how much the partial slip mechanism manifests itself in every system with a liquid-liquid interface can be resolved experimentally.

First experiment that can be proposed in this connection is the following. Consider a system of two immiscible liquids A and B. The first experiment consists of measuring the velocity $V_A$ of a small spherical droplet of liquid A with radius $R$ in a large reservoir filled with liquid B. Then, if the motion of the droplet is determined by the condition of partial slip, the slip length, according to (75), is determined by:

$$\lambda_B = \frac{(\rho_B - \rho_A)gR^2 - 3\eta_B V_A}{6(\rho_B - \rho_A)gR} - R\left(\frac{\eta_B}{3\eta_A} + \frac{1}{2}\right). \tag{109}$$

In the second experiment, the liquids A and B are swapped, so that the velocity $V_B$ of a droplet of liquid B of the same radius $R$ is measured in a large reservoir filled with liquid A. Then, instead of (109), we can write

$$\lambda_A = \frac{(\rho_A - \rho_B)gR^2 - 3\eta_A V_B}{6(\rho_A - \rho_B)gR} - R\left(\frac{\eta_A}{3\eta_B} + \frac{1}{2}\right). \tag{110}$$

If the partial slip mechanism really operates in the liquid-liquid system under consideration, relation (16) must be satisfied, namely:

$$\frac{\lambda_B}{\lambda_A} = -\frac{\eta_B}{\eta_A}. \tag{111}$$

If there is a deviation from Eq. (111), that indicates a noticeable influence of other mechanisms of interaction between liquids on the interface.

## 8. CONCLUSIONS

It is currently accepted that the HRE should describe the system under consideration well. The existing deviations, according to Levich [3] and many other researchers [23-24], should be associated with the difficulty of ensuring sufficient purity of liquids, and unaccounted surfactants distort the interpretation of experimental result. This paper shows that there is another previously unaccounted possibility for interpreting such deviations, related to the mechanism of partial slip at the liquid-liquid interface based on generalized Navier boundary condition [14-15]. Previously, the mechanism of partial slip was discussed only at the liquid-solid interface. But, given that a solid can be considered as a liquid with infinite viscosity, we transfer here the idea of partial slip to the liquid-liquid interface.

A solution to the problem of low-Reynolds-number flow around a spherical liquid droplet moving in another liquid is proposed. We obtained a generalization of the HRE to the case of non-zero slip length at the liquid-liquid interface, Eq.(75). A particular version of this solution is also considered, namely, a model in which it is assumed that all components of the viscous stress tensor, including diagonal components $\sigma_{\theta\theta}$ and $\sigma_{\phi\phi}$, are continuous at the interface (Appendix).

Using the proposed model, an experiment is considered in which the velocity of falling of a spherical drop of silicate oil in the castor oil is investigated. It is shown that the partial slip model gives a deviation from the experiment at the level of the deviation that in this case occurs for the Hadamard-Rybczynski model. In this experiment, the partial slip model predicts a liquid flow velocity inside the droplet that is an order of magnitude greater than that required by the Hadamard-Rybczynski theory (Fig. 3-4). It could be tried to be detected in an additional experiment that allows measuring the flow velocity into the droplet.

If the experiment does not allow to make a choice in favor of one model or another, taking into account that it is quite difficult to ensure the purity of the contacting liquids, the existing result apparently indicates the multiplicity of interacting mechanisms that operate in real systems.

In general, it should be noted that the hydrodynamic approach (Navier-Stokes equations) is based primarily on viscosity. In this sense, both the Boussinesq approach and the approach with a non-zero slip length are more natural precisely within the framework of that approach. At the same time, of course, one cannot deny the existence in nature of others, more specific mechanisms that are associated with the non-uniform environment of the liquid-liquid interface due to presence of surfactants [3,6,24].

The field of application of the generalized HRE given by Eq.(75) is the motion of hydrophobic (lipophilic) liquids in water and vice versa, i.e. the description of stratification and sedimentation of aqueous emulsions or vice versa – water droplets in oils, etc. Presumably, the best applicability of this equation should be expected for the interface of hydrophobic liquid and hydrophilic one (water – hydrocarbons, water – higher alcohols, in general: aqueous emulsions, water – lipophilic organic liquids and oils, etc.). These are quite important emulsions in practical terms, for example, for the oil industry and medicine. The use of such a parameter as slip length allows the experiment and theory to be reconciled.

**APPENDIX A: MODEL WITH CONTINUOUS VISCOUS STRESS TENSOR**

Conditions (91)-(92) can be rewritten in a form similar to Eq. (13):

$$-p + 2\eta\left(\frac{1}{R}\frac{\partial V_\theta}{\partial \theta} + \frac{V_r}{R}\right) + p' - 2\eta'\left(\frac{1}{R}\frac{\partial V_\theta'}{\partial \theta} + \frac{V_r'}{R}\right) = p_0 - p_0' = (\rho - \rho')gR\cos\theta, \quad (A1)$$

$$-p + 2\eta\left(\frac{V_r}{R} + \frac{V_\theta \cot\theta}{R}\right) + p' - 2\eta'\left(\frac{V_r'}{R} + \frac{V_\theta' \cot\theta}{R}\right) = p_0 - p_0' = (\rho - \rho')gR\cos\theta. \quad (A2)$$

Subtracting Eq. (A2) from Eq. (A1), we obtain:

$$2\eta\left(\frac{1}{R}\frac{\partial V_\theta}{\partial\theta}+\frac{V_r}{R}\right)-2\eta\left(\frac{V_r}{R}+\frac{V_\theta\cot\theta}{R}\right)-2\eta'\left(\frac{1}{R}\frac{\partial V_\theta'}{\partial\theta}+\frac{V_r'}{R}\right)+2\eta'\left(\frac{V_r'}{R}+\frac{V_\theta'\cot\theta}{R}\right)=0 \quad (A3)$$

or, after transformations:

$$\eta\left(\frac{\partial V_\theta}{\partial\theta}-V_\theta\cot\theta\right)=\eta'\left(\frac{\partial V_\theta'}{\partial\theta}-V_\theta'\cot\theta\right). \quad (A4)$$

Similarly to the derivation of the Hadamard–Rybczynski formula, we use the general solution presented in Table 1, taking into account that the boundary conditions require solutions containing $\cos\theta$ and $\sin\theta$, i.e. corresponding to $l=1$. For the internal problem we have solutions given by Eqs. (17)-(19), and for the external problem we have Eqs. (20)-(22).

Condition (A4) within the framework of an axisymmetric problem can be transformed to the form [25]:

$$\frac{\partial P_l^1}{\partial\theta}-P_l^1\cot\theta=-\sin\theta\frac{\partial P_l^1}{\partial\cos\theta}-\frac{\sin\theta}{\cos\theta}P_l^1=-\frac{1}{2}l(l+1)P_l+\frac{1}{2}P_l^2+\frac{1}{2}P_l^2+\frac{1}{2}l(l+1)P_l=P_l^2 \quad (A5)$$

In particular, for $l=1$ we have $P_1^2\equiv 0$, so that conditions (A1) and (A2) identically coincide.

The external fluid must satisfy the conditions at infinity. In this case, it is necessary to use the solution of the internal problem given by Eqs. (17)-(19) for $a=0$. Indeed, this will allow us to satisfy conditions (5)-(6). Substituting Eqs. (17)-(19) for $a=0$ into Eqs. (5)-(6) gives

$$2c=V_0. \quad (A6)$$

Adding here the solution of the external problem, we obtain that the external fluid is described by the equations

$$V_r=\left\{V_0-b\frac{R}{r}-2d\left(\frac{R}{r}\right)^3\right\}\cos\theta, \quad (A7)$$

$$V_\theta=\left\{\frac{bR}{2r}-d\left(\frac{R}{r}\right)^3-V_0\right\}\sin\theta, \quad (A8)$$

$$p=-\frac{\eta}{R}b\left(\frac{R}{r}\right)^2\cos\theta. \quad (A9)$$

In order to have bounded solutions inside the droplet, the internal fluid must be described only by the solution of the internal problem

$$V_r'=2\left\{\frac{a'}{10}\left(\frac{r}{R}\right)^2+c'\right\}\cos\theta, \quad (A10)$$

$$V_\theta'=-\left\{a'\frac{4}{10}\left(\frac{r}{R}\right)^2+2c'\right\}\sin\theta. \quad (A11)$$

$$p' = p_0' + \frac{2a'\eta'r}{R^2}\cos\theta. \qquad (A12)$$

We substitute the obtained relations into the boundary conditions on the droplet surface. For the radial component of the velocity of the external fluid (A7) we have

$$V_r(R,\theta) = (V_0 - b - 2d)\cos\theta = 0. \qquad (A13)$$

Hence

$$b = V_0 - 2d. \qquad (A14)$$

Then, substituting Eq. (A14) into Eqs. (A7)-(A9), we obtain the equations describing the external fluid

$$V_r = \left\{V_0\left(1 - \frac{R}{r}\right) + 2d\left[\frac{R}{r} - \left(\frac{R}{r}\right)^3\right]\right\}\cos\theta, \qquad (A15)$$

$$V_\theta = \left\{V_0\left(\frac{R}{2r} - 1\right) - d\left[\frac{R}{r} + \left(\frac{R}{r}\right)^3\right]\right\}\sin\theta, \qquad (A16)$$

$$p = -\frac{\eta}{R}(V_0 - 2d)\left(\frac{R}{r}\right)^2\cos\theta. \qquad (A17)$$

Similarly, for the radial component of the velocity of the internal fluid (A10) we have

$$V_r'(R,\theta) = 2\left\{\frac{a'}{10} + c'\right\}\cos\theta = 0, \qquad (A18)$$

Hence

$$a' = -10c'. \qquad (A19)$$

Then the system of equations for the internal fluid can be rewritten as

$$V_r' = 2c'\left\{1 - \left(\frac{r}{R}\right)^2\right\}\cos\theta, \qquad (A20)$$

$$V_\theta' = 2c'\left\{2\left(\frac{r}{R}\right)^2 - 1\right\}\sin\theta. \qquad (A21)$$

$$p' = -\frac{20c'\eta'r}{R^2}\cos\theta. \qquad (A22)$$

Conditions Eqs. (7)-(8) imposed in Eq. (A15) and Eq. (A20) on the surface of the drop give

$$\left.\frac{\partial V_r}{\partial \theta}\right|_{r=R} = \left.\frac{\partial V_r'}{\partial \theta}\right|_{r=R} = 0. \qquad (A23)$$

Therefore, the condition of stitching stresses $\sigma_{r\theta}$ giving by Eq. (12) is simplified

$$\eta\left(\left.\frac{\partial V_\theta}{\partial r}\right|_{r=R} - \frac{V_\theta(R)}{R}\right) = \eta'\left(\left.\frac{\partial V_\theta'}{\partial r}\right|_{r=R} - \frac{V_\theta'(R)}{R}\right). \qquad (A24)$$

Next, we have

$$V_\theta(R) = -\left\{\frac{1}{2}V_0 + 2d\right\}\sin\theta, \qquad (A25)$$

$$\frac{\partial V_\theta}{\partial r} = \left\{-V_0\left(\frac{R}{2r^2}\right) + d\left[\frac{R}{r^2} + \frac{3}{R}\left(\frac{R}{r}\right)^4\right]\right\}\sin\theta, \qquad (A26)$$

$$\left.\frac{\partial V_r}{\partial \theta}\right|_{r=R} = \left\{-V_0\frac{1}{2R} + d\frac{4}{R}\right\}\sin\theta = \frac{8d - V_0}{2R}\sin\theta, \qquad (A27)$$

$$\left.\frac{\partial V_r}{\partial \theta}\right|_{r=R} - \frac{V_\theta(R)}{R} = \frac{8d - V_0}{2R}\sin\theta + \frac{1}{R}\left\{\frac{1}{2}V_0 + 2d\right\}\sin\theta = \frac{6d}{R}\sin\theta. \qquad (A28)$$

Similarly, we find

$$V_\theta'(R) = 2c'\sin\theta, \qquad (A29)$$

$$\frac{\partial V_\theta'}{\partial r} = 2c'\frac{1}{R}\left\{4\frac{r}{R}\right\}\sin\theta, \qquad (A30)$$

$$\left.\frac{\partial V_\theta'}{\partial r}\right|_{r=R} = 8c'\frac{1}{R}\sin\theta, \qquad (A31)$$

$$\left.\frac{\partial V_\theta'}{\partial r}\right|_{r=R} - \frac{V_\theta'(R)}{R} = \frac{6c'}{R}\sin\theta, \qquad (A32)$$

Substituting Eq. (A28) and Eq. (A32) into Eq. (A24), we obtain

$$c' = \frac{\eta}{\eta'}d. \qquad (A33)$$

Then

$$V_r' = 2\frac{\eta}{\eta'}d\left\{1 - \left(\frac{r}{R}\right)^2\right\}\cos\theta, \qquad (A34)$$

$$V_\theta' = 2\frac{\eta}{\eta'}d\left\{2\left(\frac{r}{R}\right)^2 - 1\right\}\sin\theta. \qquad (A35)$$

$$p' = -\frac{\eta}{\eta'}d\frac{20\eta'r}{R^2}\cos\theta = -d\frac{20\eta r}{R^2}\cos\theta. \qquad (A36)$$

Let's calculate the expression

$$-p(R) + 2\eta\left.\frac{\partial V_r}{\partial r}\right|_R = \frac{3\eta}{R}(V_0 + 2d)\cos\theta. \qquad (A37)$$

Similarly, we find

$$-p'(R) + 2\eta'\frac{\partial V_r'}{\partial r}\bigg|_{r=R} = d\frac{20\eta r}{R^2}\cos\theta - \frac{8\eta'\eta}{\eta'R^2}d\cos\theta = \frac{12d\eta}{R}\cos\theta. \qquad (A38)$$

Substituting Eq. (A37) and Eq. (A38) into the boundary condition Eq. (13), we obtain

$$V_0 = 2d + \frac{(\rho - \rho')gR^2}{3\eta} \qquad (A39)$$

The remaining boundary condition Eq. (A2), taking into account conditions (7)-(8), has the form

$$-p(R) + 2\eta\frac{V_\theta(R)\cot\theta}{R} + p'(R) - 2\eta'\frac{V_\theta'(R)\cot\theta}{R} = (\rho - \rho')gR\cos\theta. \qquad (A40)$$

Hence

$$d = -\frac{(\rho - \rho')gR^2}{30\eta}. \qquad (A41)$$

Substituting Eq. (A41) into Eq. (A39), one can obtain

$$V_0 = \frac{4(\rho - \rho')gR^2}{15\eta}. \qquad (A42)$$

Thus, replacing the condition for stitching the velocities of two liquids, internal and external, allows us to take into account the remaining unconsidered conditions for stitching stresses. The velocity (A42) exceeds the velocity of the solid sphere, determined by the Stokes drag force Eq. (79), by $6/5$ times. At the same time, there is no dependence on the viscosity of the liquid into droplet, $\eta'$, which qualitatively distinguishes formula (A42) from the Hadamard–Rybczynski Eq. (76).

The remaining undefined coefficients in Eqs. (A6), (A14), (A33) and (A19) take the form

$$c = \frac{2(\rho - \rho')gR^2}{15\eta}, \; b = \frac{(\rho - \rho')gR^2}{3\eta}, \; c' = -\frac{(\rho - \rho')gR^2}{30\eta'}, \; a' = \frac{(\rho - \rho')gR^2}{3\eta'}. \qquad (A43)$$

Let us write out the solutions for the external and internal liquids by substituting Eq. (A41) in Eqs. (A15)-(A17) and Eqs. (A34)-(A36), respectively.

1) The external liquid

$$V_r = \left\{4 - \frac{5R}{r} + \left(\frac{R}{r}\right)^3\right\}\frac{(\rho - \rho')gR^2}{15\eta}\cos\theta, \qquad (A44)$$

$$V_\theta = \left\{5\frac{R}{r} + \left(\frac{R}{r}\right)^3 - 8\right\}\frac{(\rho - \rho')gR^2}{30\eta}\sin\theta, \qquad (A45)$$

$$p = -\frac{(\rho - \rho')gR}{3}\left(\frac{R}{r}\right)^2\cos\theta. \qquad (A46)$$

Stream function is (Table 1):

$$\Psi = R^2 \sin\theta \left\{ \frac{b}{2}\left(\frac{R}{r}\right)^{-1} + d\left(\frac{R}{r}\right) - c\left(\frac{r}{R}\right)^2 \right\} P_1^1 \quad (A47)$$

or

$$\Psi = \frac{(\rho-\rho')gR^4}{30\eta} \sin^2\theta \left\{ \frac{R}{r} - \frac{5r}{R} + 4\left(\frac{r}{R}\right)^2 \right\}. \quad (A48)$$

2) The internal liquid

$$V_r' = \frac{(\rho-\rho')gR^2}{15\eta'}\left\{\left(\frac{r}{R}\right)^2 - 1\right\}\cos\theta, \quad (A49)$$

$$V_\theta' = \frac{(\rho-\rho')gR^2}{15\eta'}\left\{1 - 2\left(\frac{r}{R}\right)^2\right\}\sin\theta. \quad (A50)$$

$$p' = \frac{2(\rho-\rho')gr}{3}\cos\theta. \quad (A51)$$

Stream function is (Table 1):

$$\Psi' = -R^2 \sin\theta \left\{ \frac{a'}{10}\left(\frac{r}{R}\right)^4 + c'\left(\frac{r}{R}\right)^2 \right\} P_1^1. \quad (A52)$$

or

$$\Psi' = \frac{(\rho-\rho')gR^4}{30\eta'}\sin^2\theta\left\{\left(\frac{r}{R}\right)^4 - \left(\frac{r}{R}\right)^2\right\}. \quad (A53)$$

On the surface of a drop, the tangential components of the velocities are determined by the formulae

$$V_\theta(R,\theta) = -\frac{(\rho-\rho')gR^2}{15\eta}\sin\theta, \quad V_\theta'(R,\theta) = -\frac{(\rho-\rho')gR^2}{15\eta'}\sin\theta, \quad (A54)$$

Then relative slip velocity is:

$$\Delta V_\theta(R,\theta) = V_\theta(R,\theta) - V_\theta'(R,\theta) = \frac{(\rho-\rho')gR^2}{15}\left(\frac{1}{\eta'} - \frac{1}{\eta}\right)\sin\theta. \quad (A55)$$

If the viscosities of the two liquids are the same, the slip velocity given by Eq. (A55) vanishes, so we return to the no-slip boundary condition, $V_\theta(R,\theta) = V_\theta'(R,\theta)$, that corresponds to the Hadamard and Rybczynski model.

Similar to the Navier boundary condition used to describe the slip of a liquid on a solid surface, we can introduce an effective slip length for a liquid drop in an external liquid:

$$\sigma_{\theta r}(\theta) = \eta\left(\frac{1}{r}\frac{\partial V_r}{\partial \theta} + \frac{\partial V_\theta}{\partial r} - \frac{V_\theta}{r}\right)_{r=R} = \frac{\eta}{\lambda}(V_\theta(R,\theta) - V_\theta'(R,\theta)). \quad (A56)$$

According to equation (A24), we have

$$\sigma_{\theta r}(\theta) = \sigma_{\theta r}{}'(\theta) = \eta\left(\left.\frac{\partial V_\theta}{\partial r}\right|_{r=R} - \frac{V_\theta(R)}{R}\right) = \eta'\left(\left.\frac{\partial V_\theta{}'}{\partial r}\right|_{r=R} - \frac{V_\theta{}'(R)}{R}\right), \tag{A57}$$

where

$$\left.\frac{\partial V_\theta}{\partial r}\right|_{r=R} = -4\frac{(\rho-\rho')gR}{15\eta}\sin\theta. \tag{A58}$$

Let us calculate the slip length $\lambda$ from the equation

$$\sigma_{\theta r}(\theta) = \eta\left(\left.\frac{\partial V_\theta}{\partial r}\right|_{r=R} - \frac{V_\theta(R,\theta)}{R}\right) = \frac{\eta}{\lambda}(V_\theta(R,\theta) - V_\theta{}'(R,\theta)), \tag{A59}$$

$$-\frac{(\rho-\rho')gR}{5}\sin\theta = \frac{\eta}{\lambda}\frac{(\rho-\rho')gR^2}{15}\left(\frac{1}{\eta'}-\frac{1}{\eta}\right)\sin\theta. \tag{A60}$$

or

$$\lambda = \frac{R}{3}\left(1 - \frac{\eta}{\eta'}\right). \tag{A61}$$

Similarly, for the internal fluid, the slip length $\lambda'$ is given by the relation

$$\sigma_{\theta r}{}'(\theta) = \eta'\left(\left.\frac{\partial V_\theta{}'}{\partial r}\right|_{r=R} - \frac{V_\theta{}'(R,\theta)}{R}\right) = \frac{\eta'}{\lambda'}(V_\theta{}'(R,\theta) - V_\theta(R,\theta)), \tag{A62}$$

hence

$$\lambda' = \frac{R}{3}\left(1 - \frac{\eta'}{\eta}\right). \tag{A63}$$

---